\documentclass[12pt]{article}
\pdfoutput=1
\usepackage{amsmath}
\usepackage{amssymb}
\usepackage{graphicx}
\usepackage{color}
\newlength{\abstractwidth}
\usepackage[T1]{fontenc}
\usepackage[utf8]{inputenc}
\usepackage{graphicx}
\usepackage{bm}
\usepackage{xcolor}
\usepackage{slashed}
\usepackage{cite}
\usepackage[hidelinks]{hyperref}

\newcommand{\be}{\begin{equation}}
\newcommand{\ee}{\end{equation}}
\newcommand{\bea}[1]{\begin{align*}#1\end{align*}}
\newcommand\eqn{\addtocounter{equation}{1}\tag{\theequation}}

\def\k{\mathbf{k}}

\def\D{\Delta}
\def\Z{\mathcal{Z}_{in/out}}

\DeclareMathOperator{\sech}{sech}
\DeclareMathOperator{\re}{Re}
\DeclareMathOperator{\im}{Im}
\DeclareMathOperator{\Tr}{Tr}

\begin{document}

\begin{titlepage}
\bigskip
\bigskip\bigskip 
\bigskip

\begin{center}
{\Large \bf{Scalar and Spinor Effective Actions in Global de Sitter}}
 \bigskip
{\Large \bf { }} 
\bigskip
\bigskip
\end{center}

\begin{center}

\bf {Jiaqi Jiang}
\bigskip \rm
\bigskip \rm

\bigskip
\bigskip
Jadwin Hall, Princeton University,  Princeton, NJ 08540, USA

\end{center}
\bigskip\bigskip
\begin{abstract}
\medskip
\noindent
In this paper, we compute the effective action of both a scalar field and a Dirac spinor field in the global de Sitter space of any dimension $d$ using the in-/out-state formalism. We show that there is particle production in even dimensions for both scalar field and spinor field. The in-out vacuum amplitude $\Z$ is divergent at late times. By using dimensional regularization, we extract the finite part of $\log\Z$ for $d$ even and the logarithmically divergent part of $\log\Z$ for $d$ odd.  We also find that the regularized in-out vacuum amplitude equals the ratio of determinants associated with different quantizations in $AdS_d$ upon the identification of certain parameters in the two theories.
\end{abstract}
\bigskip \bigskip \bigskip 
 
\end{titlepage}

\tableofcontents

\section{Introduction}
\label{sec:intro}
There has been always interest in the study of quantum fields in a de Sitter (dS) background as it is a good approximation to the expanding universe we are currently in. Just like anti-de Sitter (AdS) space, the dS space also has a maximal isometry group. As a result, the canonical quantization of free fields is possible in dS space and the effective action can be computed analytically \cite{Candelas:1975du,Dowker:1975xj,Mottola:1984ar,Allen:1985ux,Bousso:2001mw,Fukuma:2013mx,Joung:2006gj,Akhmedov:2019esv}. It has been known for a long time that dS space has a one-parameter family of vacua invariant under the dS isometry group \cite{Mottola:1984ar,Allen:1985ux,Bousso:2001mw}. Thus, the calculation of the effective action requires the specification of the vacuum states. One of the guiding principles for selection of the vacuum states is the composition principle proposed by Polyakov \cite{Polyakov:2007mm}. While the results in \cite{Candelas:1975du,Dowker:1975xj} are calculated using the in-/in-state (Schwinger-Keldysh) formalism with the Bunch-Davies vacuum, it is the in-/out-state formalism that complies with the composition principle. 

Although the effective action of a scalar field in the global patch of $dS_d$ space in the in-/out-state formalism has been computed in \cite{Kim:2010cb,Akhmedov:2019esv}, the authors have been focusing on the imaginary part of the effective action. Besides, as far as we are aware there has been no attempts in computing such effective action of a Dirac spinor field in dS space of arbitrary dimension. In this paper we use the in-/out-state formalism to compute the effective action of both a massive scalar field and a massive Dirac spinor field in the global patch of dS space of any dimension $d$ given by the metric $g_{\mu\nu}$ as
\be\label{eq:metric_dS}
ds^2=-dt^2+H^{-2}\cosh^2(Ht)d\Omega_{d-1}^2,
\ee
where $H$ is related to the scalar curvature $R$ by $R=d(d-1)H^2$ and $d\Omega^2_{d-1}$ denotes the metric of a unit $d-1$ sphere. We will set $H=1$ throughout the rest of the paper. The dependence on $H$ can be restored via dimensional analysis. 

Due to the infinite volume of the dS space, the effective action has IR divergence. However in odd dimensions the divergent effective action is pure real, while in even dimensions it also has an imaginary part. Since the divergence of the imaginary part is simply proportional to the volume of the dS space, it can be interpreted as $\mathcal{P}\times V_{dS}$ where $\mathcal{P}$ is the particle production rate per volume and $V_{dS}$ is the volume of the dS space \cite{Mottola:1984ar,Anderson:2013ila}. In contrast, the divergence of the real part is more complicated and we need to use a more sophiscated regularization method. Nevertheless, one can choose some cut-off $T$ on the time of global dS space and evolve the system from a finite initial time $-T$ to a finite time $T$. One then defines time-dependent adiabatic vacuum states which are the instantaneos ground state of the time-dependent Hamiltonian \cite{Mottola:1984ar,Anderson:2013ila,Fukuma:2013mx}. These adiabatic vacuum states interpolate between the in-state $|0,\text{in}\rangle$  and out-state $|0,\text{in}\rangle$ which we define in section \ref{sec:in out form}. The $T\to\infty$ limit is taken at the end. With this cut-off $T$, we expect the structure of the effective action $W$ to be as
\be\label{eq:div_structure_even}
W\sim c_1\int d^{d-1}x\!\sqrt{\tilde{g}}+c_2\int d^{d-1}x\!\sqrt{\tilde{g}}\mathcal{\tilde{R}}+\dots +W_{finite}
\ee
for $d$ even, and as
\be\label{eq:div_structure_odd}
W\sim c_1\int d^{d-1}x\!\sqrt{\tilde{g}}+c_2\int d^{d-1}x\!\sqrt{\tilde{g}}\mathcal{\tilde{R}}+\dots +c \log \tilde{R}+W_{finite}
\ee
for $d$ odd. In \eqref{eq:div_structure_even} and \eqref{eq:div_structure_odd}, $\tilde{g}$, $\tilde{\mathcal{R}}$, and $\tilde{R}$ are the determinant of the metric, the scalar curvature, and the radius of the spatial slice at the cut-off time $T$ respectively. For odd $d$, due to the presence of the logarithmically divergent term, the finite piece $W_{finite}$ is ambiguous. Using dimensional regularization, we will compute the finite term for even $d$ and the coefficient of the logarithmically divergent term for odd $d$. It turns out that the regularized in-out vacuum amplitude has the same expression as the ratio of determinants associated with different quantizations in AdS space. The computation of such ratio of determinants is related to the double-trace deformation in AdS/CFT correspondence\cite{Gubser:2002zh,Gubser:2002vv,Diaz:2007an,Allais:2010qq,Aros:2011iz}.

The paper is organized as follows. In section \ref{sec:in out form}, we briefly review the calculation of the effective action using the in-/out-state formalism. Section \ref{sec:massive scalar} contains the calculation for a massive real scalar field and section \ref{sec:massive dirac} contains the calculation for a massive Dirac spinor field. In section \ref{sec:particle production} we focus on the imaginary part of the effective action and obtain the particle production rate in even dimensions. In section \ref{sec:effaction dim_reg} we use dimensional regularization to obtain a closed-form expression for the vacuum amplitude. For even $d$ we get a finite answer, while for odd $d$ we compute the coefficient of the logarithmically divergent term. We further show that this expression equals to the ratio of determinants associated with different quantizations in AdS space in both scalar and spinor cases. In the concluding section we briefly discuss the possible connections between these two qunatities.

\section{In-/out- state formalism}
\label{sec:in out form}
In this section, we briefly review the in-/out- state formalism for calculating the effective action \cite{DeWitt:1975ys,Nikishov:2002ez,Kim:2008yt}. The effective action $W$ is defined by the scattering amplitude
\be
\mathcal{Z}_{in/out}=e^{iW}=\langle 0,\text{out}|0,\text{in}\rangle,\quad W=\int d^d x \sqrt{-g}\,\mathcal{L}_{eff},
\ee
where $|0,\text{in}\rangle$ and $|0,\text{out}\rangle$ are the in-going and the out-going vacuum states respectively. 
Specifically, for a general massive field $\Phi$, we can expand it in terms of either in-going modes or out-going modes over the set of quantum numbers $\lambda$ as
\be
\Phi=\sum_{\lambda}a_{\lambda,in}\Phi_{\lambda+}+b_{\lambda,in}^\dagger\Phi_{\lambda-}=\sum_{\lambda}a_{\lambda,out}{\Phi_{\lambda}}^++b_{\lambda,out}^\dagger{\Phi_{\lambda}}^-.
\ee
Here $\Phi_{\lambda+}$ and $\Phi_{\lambda-}$ are the positive-frequency and negative-frequency in-going modes while ${\Phi_{\lambda}}^+$ and ${\Phi_{\lambda}}^-$ are the positive-frequency and negative-frequency out-going modes. In terms of the asymptotic behavior, we have 
\begin{align}
&\Phi_{\lambda\pm}\sim e^{\mp i\mu t} \quad \text{ as } t\rightarrow -\infty,\\
&{\Phi_{\lambda}}^\pm\sim e^{\mp i\mu t} \quad \text{ as } t\rightarrow +\infty,
\end{align}
where $\mu$ is some effective mass.
Then $|0,\text{in}\rangle$ is the state annihilated by $a_{\k,in}$ and $b_{\lambda,in}$ for each $\lambda$ while $|0,\text{out}\rangle$ is the state annihilated  by $a_{\lambda,out}$ and $b_{\lambda,out}$. 

For a scalar field $\Phi$, the in-going modes $\Phi_{\lambda\pm}$ and the out-going modes ${\Phi_{\lambda}}^\pm$ are related by the Bogoliubov transformation \cite{Mottola:1984ar,Nikishov:2002ez}:
\bea{\label{eq:bog_scalar}
\Phi_{\lambda\,+}&=\mu_{\lambda}\,\Phi_\lambda^{\,\, +}+\nu_{\lambda}\,\Phi_\lambda^{\,\, -},\\
\Phi_{\lambda\,-}&=\nu_{\lambda}^*\,\Phi_\lambda^{\,\, +}+\mu_{\lambda}^*\,\Phi_\lambda^{\,\, -},
\eqn
}
with the coefficients $\mu_{\lambda}$ and $\nu_{\lambda}$ satisfying the Bogoliubov relation
\be
|\mu_\lambda|^2-|\nu_\lambda|^2=1,
\ee
as required by the commutation relations for bosons. We emphasize here that we have assumed the transformation \eqref{eq:bog_scalar} is diagonal in $\lambda$, which is true in the case studied. Smilarly for a spinor field $\Psi$, the in-going modes $\Psi_{\lambda\pm}$ and the out-going modes $\Psi_{\lambda}^{\,\pm}$ are related by the transformation \cite{Nikishov:2002ez}
\bea{\label{eq:bog_spinor}
\Psi_{\lambda\,+}&=\mu_{\lambda}\,\Psi_\lambda^{\,\, +}+\nu_{\lambda}\,\Psi_\lambda^{\,\, -},\\
\Psi_{\lambda\,-}&=-\nu_{\lambda}^*\,\Psi_\lambda^{\,\, +}+\mu_{\lambda}^*\,\Psi_\lambda^{\,\, -}.
\eqn
}
The coefficients $\mu_\lambda$ and $\nu_\lambda$ now satisfy the relation
\be
|\mu_\lambda|^2+|\nu_\lambda|^2=1,
\ee
as required by the commutation relations for fermions. Again we have assumed the transformation \eqref{eq:bog_spinor} is diagonal in $\lambda$, as this is the case of interest.

In the in-/out-state formalism, the exact one-loop effective action $W$ can be then expressed in terms of the Bogoliubov coefficients $\mu_\k$ as
\be\label{eq:form_effact}
W=\int d^d x \sqrt{-g}\,\mathcal{L}_{eff}=\sum_{\lambda} d(\lambda)W_\lambda,
\ee
where $d(\lambda)$ is the degeneracy and $W_\lambda$ is given by \footnote{For a real massive scalar, there is a factor of $\frac{1}{2}$ as a complex scalar field can be viewed as two real scalar fields.}
\be
W_\lambda=
\begin{cases}
i\ln \mu_\lambda^* &\text{ (complex scalar)},\\
-i\ln \mu_\lambda^* &\text{ (Dirac spinor)}.
\end{cases}
\ee

\section{Massive scalar field in global de Sitter}
\label{sec:massive scalar}
In this section, we consider a real scalar field $\Phi$ with mass $M$ in the global patch of $dS_d$ space. We briefly review the calculation here as it has been done in various papers \cite{Mottola:1984ar,Bousso:2001mw,Joung:2006gj,Kim:2010cb,Fukuma:2013mx,Akhmedov:2019esv}. The action of the scalar field is
\be
-\frac{1}{2}\int d^dx \sqrt{-g}\biggl( \partial_\mu\Phi\partial^\mu\Phi+M^2\Phi^2\biggr).
\ee
From the action, we have that the Klein-Gordon equation for $\Phi$ is
\be
\left[-\frac{1}{\cosh^{d-1}(t)}\partial_t\left(\cosh^{d-1}(t)\partial_t\right)+\frac{1}{\cosh^2(t)}\nabla^2_{\Omega_{d-1}}-M^2\right]\Phi=0
\ee
where $\nabla^2_{\Omega_{d-1}}$ is the Laplacian on the unit $d-1$ sphere. To solve the equation, we expand $\Phi$ using the real spherical harmonics $Y_l(\Omega_{d-1})$ which satisfies
\be
\nabla^2_{\Omega_{d-1}}Y_l(\Omega_{d-1})=-l(l+d-2)\,Y_l(\Omega_{d-1}),
\ee
with degeneracies:
\be
D_{(d-1)}(l)=\frac{(l+d-3)!}{l!\,(d-2)!}(2l+d-2),\quad (l=0,1,\dots).
\ee
Using the expansion $\Phi(t,\Omega_{d-1})=\sum_l\,\phi_l(t)Y_l(\Omega_{d-1})$, we find that for each mode $l$ the function $\phi_l(t)$ satisfies the equation:
\be\label{eq:KG-scalar}
\left(\partial_t^2+(d-1)\tanh(t)\partial_t+\frac{l(l+d-2)}{\cosh^2(t)}+M^2\right)\phi_l(t)=0.
\ee
Following \cite{Mottola:1984ar,Bousso:2001mw,Kim:2010cb}, we could write the two independent solutions as either
\bea{\label{eq:sol_scalar_1}
\phi_l^{\,\,(\pm)}(t)=&\cosh^l(t)\exp\left[\left(-l-\frac{d-1}{2}\mp i \mu\right)t\right]\\
&F\left(l+\frac{d-1}{2},l+\frac{d-1}{2}\pm i\mu,1\pm i\mu,-e^{-2t}\right),
\eqn
}
or
\bea{\label{eq:sol_scalar_2}
\phi_{l\,(\pm)}(t)=&\cosh^l(t)\exp\left[\left(l+\frac{d-1}{2}\mp i \mu\right)t\right]\\
&F\left(l+\frac{d-1}{2},l+\frac{d-1}{2}\mp i\mu,1\mp i\mu,-e^{2t}\right),
\eqn    
}
where $F$ is the hypergeometric function $_2F_1$ and $\mu=\sqrt{M^2-\frac{(d-1)^2}{4}}$. Here we have restricted our attention to the case $M^2>(d-1)^2/4$, where the solution oscillates at the past and the future infinity
\footnote{The solutions for $M^2<(d-1)^2/4$ can be obtained by analytic continuation in $\gamma$. In this case, the modes do not oscillate and can be interpreted as the source of the operator in the dual boundary CFT \cite{Strominger:2001pn}.}.  The two solutions \eqref{eq:sol_scalar_1} have the asymptotic behaviors as
\be
\phi_l^{\,\,(\pm)}(t)\rightarrow \exp\left(-\frac{d-1}{2}t\mp i\mu t\right) \text{ as } t\rightarrow +\infty,
\ee
while the two solutions \eqref{eq:sol_scalar_2} have the asymptotic behaviors as
\be
\phi_{l\,(\pm)}(t)\rightarrow \exp\left(\frac{d-1}{2}t\mp i\mu t\right) \text{ as } t\rightarrow-\infty.
\ee
Therefore, we could identify the two solutions \eqref{eq:sol_scalar_1} as the positive/negative-frequency out-modes while the two solutions \eqref{eq:sol_scalar_2} as the positive/negative-frequency in-modes. The in-modes and the out-modes are related by the Bogoliubov transformation \eqref{eq:bog_scalar} as
\bea{
\phi_{l\,+}&=\mu_l\,\phi_l^{\,\, +}+\nu_l\,\phi_l^{\,\, -},\\
\phi_{l\,-}&=\nu_l^*\,\phi_l^{\,\, +}+\mu_l^*\,\phi_l^{\,\, -},
\eqn
}
where we have suppressed the dependence on $t$.
Using the transformation formula for the hypergeometric function \cite{NIST:DLMF}, we find the Bogoliubov coefficients to be
\bea{
\mu_{\,l}&=\frac{\Gamma(1-i\mu)\Gamma(-i\mu)}{\Gamma(l+\frac{d-1}{2}-i\mu)\Gamma(-l-\frac{d-3}{2}-i\mu)}, \quad (l=0,1,\dots),\\
\nu_{\,l}&=\frac{i\cos(l\pi+\frac{d}{2}\pi)}{\sinh(\pi\mu)},
\eqn
}
with degeneracies $D_{(d-1)}(l)$. One can check that these coefficients indeed obey the relation
\be
|\mu_l|^2-|\nu_l|^2=1,
\ee
as required by the commutation rules. In particular, $\nu_l=0$ when $d$ is odd, which implies that $|0,\text{in}\rangle$ and $|0,\text{out}\rangle$ define the same state.

\section{Massive Dirac spinor field in global de Sitter}
\label{sec:massive dirac}
In this section, we consider a massive Dirac spinor field $\Psi$ with mass $M$ in the global patch of $dS_d$ space. The action for the Dirac field $\Psi$ is
\be
-\int d^dx \bar{\Psi}\bigl(\slashed{\nabla}+M\bigr)\Psi,
\ee
where $\slashed{\nabla}\equiv\gamma^a(\mathbf{e}_a)^\mu\nabla_\mu$. This leads to the Dirac equation
\be\label{eq:Dirac_general}
\gamma^a(\mathbf{e}_a)^\mu\left(\partial_\mu-\frac{1}{8}\omega_{\mu bc}\left[\gamma^b,\gamma^c\right]\right)\Psi+M\Psi=0.
\ee
Here $\gamma^a$ for $(a=0,\dots,d-1)$ are the gamma matrices which satisfy the Dirac algebra
\be\label{eq:id_gamma}
\gamma^a\gamma^b+\gamma^b\gamma^a=2\eta^{ab}\mathbf{1}
\ee
with $\eta^{ab}$ of the signature $(-,+,\dots,+)$.
The $\{\mathbf{e}_a\}$ is a vielbein on $dS_d$ and the spin connection $\omega_{abc}$ is defined as
\be
\omega_{abc}=(\mathbf{e}_a)^\mu\left[\partial_\mu(\mathbf{e}_b)^\nu+\Gamma^{\nu}_{\,\,\mu\gamma}(\mathbf{e}_b)^\gamma\right](\mathbf{e}_c)_\nu,
\ee
where $\{\Gamma^{\nu}_{\,\,\mu\gamma}\}$ is the Christoffel symbols of the Levi-Civita connection for the metric \eqref{eq:metric_dS}. We follow the method used in \cite{Camporesi:1995fb} to solve the Dirac equation. If we let $\{\tilde{\mathbf{e}}_i\}$ be a vielbein on the $S^{d-1}$, then we could define $\{\mathbf{e}_a\}$ as
\be
\mathbf{e}_0=\partial_t,\quad \mathbf{e}_j=\frac{1}{\cosh(t)}\tilde{\mathbf{e}}_j, \quad (j=1,\dots d-1)
\ee
The only non-zero components of the spin connection $w_{abc}$ are
\bea{\label{eq:comp_spinconnection}
\omega_{ijk}&=\frac{1}{\cosh(t)}\tilde{\omega}_{ijk},\\
\omega_{i0k}&=-\omega_{ik0}=\tanh(t)\delta_{ik},\quad (i,j,k=1,\dots,d-1),
\eqn
}
where $\tilde{\omega}_{ijk}$ is the spin connection on $S^{d-1}$ corresponding to the frame $\{\tilde{\mathbf{e}}_i\}$.

Since the construction of the representations for Clifford algebra in even and odd dimensions is slightly different, we shall discuss the two cases separately below. Our construction of the representations of the Clifford algebra \eqref{eq:id_gamma} follows \cite{Camporesi:1995fb}.

\subsection{Even dimension}
We construct the gamma matrices satisfying \eqref{eq:id_gamma} in the following way: We let $\{\gamma^a\}$ be the set of $d$ matrices of dimension $2^{d/2}$ defined by
\be\label{eq:def_gamma_even}
\gamma^0=
\begin{pmatrix}
0 & i\mathbf{1}\\
i\mathbf{1} & 0
\end{pmatrix},
\quad 
\gamma^j=
\begin{pmatrix}
0 & i\tilde{\Gamma}^j \\
-i\tilde{\Gamma}^j & 0
\end{pmatrix},
\quad (j=1,\dots, d-1)
\ee
where $\mathbf{1}$ is the identity matrix of dimension $2^{d/2-1}$ and the $d-1$ matrices $\tilde{\Gamma}^j$ also of dimension $2^{d/2-1}$ satisfy the following Clifford algebra:
\be\label{eq:clifford_algebra}
\tilde{\Gamma}^j\tilde{\Gamma}^k+\tilde{\Gamma}^k\tilde{\Gamma}^j=2\delta^{jk}\mathbf{1},
\quad (j,k=1,\dots,d-1).
\ee
With the representations of the gamma matrices defined in \eqref{eq:def_gamma_even} and \eqref{eq:comp_spinconnection}, the Dirac equation \eqref{eq:Dirac_general} becomes
\be\label{eq:dirac_even_0}
\gamma^0\left(\partial_t+\frac{d-1}{2}\tanh(t)\right)\Psi+\frac{1}{\cosh(t)}
\begin{pmatrix}
0 & i\tilde{\slashed{\nabla}}\\
-i\tilde{\slashed{\nabla}} & 0
\end{pmatrix}\Psi
+M\Psi = 0
\ee
where $\tilde{\slashed{\nabla}}$ is the Dirac operator on $S^{d-1}$. If we represent $\Psi$ with two components given by
\be
\Psi=
\begin{pmatrix}
\psi_1\\
\psi_2
\end{pmatrix},
\ee
then the Dirac equation \eqref{eq:dirac_even_0} decomposes to the following set of equations:
\bea{\label{eq:dirac_even}
i\left(\partial_t+\frac{d-1}{2}\tanh(t)+\frac{1}{\cosh(t)}\tilde{\slashed{\nabla}}\right)\psi_2+M\psi_1 &=0,\\
i\left(\partial_t+\frac{d-1}{2}\tanh(t)-\frac{1}{\cosh(t)}\tilde{\slashed{\nabla}}\right)\psi_1+M\psi_2 &=0.
\eqn
}
By eliminating either $\psi_1$ or $\psi_2$ in \eqref{eq:dirac_even}, we obtain
\bea{\label{eq:dirac_even_psi1}
\psi_1''+(d-1)\tanh(t)\psi_1'+&\left[M^2+\frac{(d-1)^2}{4}-\frac{(d-1)(d-3)}{4\cosh^2(t)}\right]\psi_1\\
&-\sech^2(t)\tilde{\slashed{\nabla}}^2\psi_1+\frac{\tanh(t)}{\cosh(t)}\tilde{\slashed{\nabla}}\psi_1 =0,\eqn\\
\label{eq:dirac_even_psi2}
\psi_2''+(d-1)\tanh(t)\psi_2'+&\left[M^2+\frac{(d-1)^2}{4}-\frac{(d-1)(d-3)}{4\cosh^2(t)}\right]\psi_2\\
&-\sech^2(t)\tilde{\slashed{\nabla}}^2\psi_2-\frac{\tanh(t)}{\cosh(t)}\tilde{\slashed{\nabla}}\psi_2 =0,
\eqn
}
where prime denotes derivatives with respect to $t$. We only need to solve the equation for $\psi_1$ and the solution for $\psi_2$ could be obtained from \eqref{eq:dirac_even}. Using the eigenfunctions of the Dirac operator $\tilde{\slashed{\nabla}}$ on $S^{d-1}$ for $d$ even which are defined by \cite{Camporesi:1995fb}
\be\label{eq:even_spin_harm}
\tilde{\slashed{\nabla}}\chi^{(\pm)}_{lm}(\Omega_{d-1})=\pm i\left(l+\frac{d-1}{2}\right)\chi^{(\pm)}_{lm}(\Omega_{d-1}),\quad (l=0,1,\dots),
\ee
with degeneracies given by
\be
\mathcal{D}_{d-1}^{(\pm)}(l)=\frac{2^{(d-2)/2}(d+l-2)!}{l!\,(d-2)!},\quad \text{for even } d.
\ee
we can separate variables by considering the expansion
\be\label{eq:psi1_exp}
\psi_1(t,\Omega_{d-1})=\sum_{l,m} \phi_l(t)\chi^{+}_{lm}(\Omega_{d-1})+\varphi_l(t)\chi^{-}_{lm}(\Omega_{d-1}).
\ee
Inserting the expansion \eqref{eq:psi1_exp} into \eqref{eq:dirac_even_psi1} and using \eqref{eq:even_spin_harm}, we obtain the equations for $\phi_l$ and $\varphi_l$ as 
\bea{\label{eq:dirac_even_phi}
\phi_l''+&(d-1)\tanh(t)\phi_l'\\
+&\left[M^2+\frac{(d-1)^2}{4}+\frac{(d-1)(2l+1)+2l^2}{2\cosh^2(t)}+i\frac{(d-1+2l)\tanh(t)}{2\cosh(t)}\right]\phi_l=0,\eqn\\
\label{eq:dirac_even_varphi}
\varphi_l''+&(d-1)\tanh(t)\varphi_l'\\
+&\left[M^2+\frac{(d-1)^2}{4}+\frac{(d-1)(2l+1)+2l^2}{2\cosh^2(t)}-i\frac{(d-1+2l)\tanh(t)}{2\cosh(t)}\right]\varphi_l=0.
\eqn
}
To solve equations \eqref{eq:dirac_even_phi} and \eqref{eq:dirac_even_varphi}, we perform a change of variable to $z=i\sinh(t)$ and consider the following ansatz:
\bea{
\phi_l(z)&=(1+z)^{l/2}(1-z)^{(l+1)/2}g_\phi(z),\\ 
\varphi_l(z)&=(1+z)^{(l+1)/2}(1-z)^{l/2}g_\varphi(z). \eqn
}
Plugging the ansatz, we find the two independent solutions for $g_\phi$ and $g_\varphi$ as
\bea{
g_{\phi\,1}(z)=&\left(\frac{1-z}{2}\right)^{-d/2-l}\tilde{F}\left(-i M, iM,1-\frac{d}{2}-l;\frac{1-z}{2}\right), \eqn\\
g_{\phi\,2}(z)=& \tilde{F}\left(\frac{d}{2}-i M+l,\frac{d}{2}+iM+l,1+\frac{d}{2}+l;\frac{1-z}{2}\right), \eqn\\
g_{\varphi\,1}(z)=&\left(\frac{1-z}{2}\right)^{-d/2-l+1}\tilde{F}\left(1-i M, 1+iM,2-\frac{d}{2}-l;\frac{1-z}{2}\right), \eqn\\
g_{\varphi\,2}(z)=&\tilde{F}\left(\frac{d}{2}-i M+l,\frac{d}{2}+iM+l,\frac{d}{2}+l;\frac{1-z}{2}\right), \eqn\\
}
where for convenience we have defined the rescaled hypergeometric function $\tilde{F}$ as
\be
\tilde{F}\left(a,b,c;x\right)=\frac{\Gamma(c-b)}{\Gamma(1-b)\Gamma(c)}\,_2F_1(a,b,c;x).
\ee
We can use the equation \eqref{eq:dirac_even} to find the corresponding solutions for $\psi_2$ component. The general solution for $\Psi$ can be written as:
\bea{\label{eq:psi_even_gensol}
\Psi=\sum_{l,m}& C_{lm}
\begin{pmatrix}
\phi_{l\,1}(t)\chi_{lm}^+(\Omega_{d-1})\\
\\
-i\varphi_{l\, 1}(t)\chi_{lm}^+(\Omega_{d-1})
\end{pmatrix}
+D_{lm}
\begin{pmatrix}
\phi_{l\,2}(t)\chi_{lm}^+(\Omega_{d-1})\\
\\
i\varphi_{l\, 2}(t)\chi_{lm}^+(\Omega_{d-1})
\end{pmatrix}\\
&+\sum_{l,m}C'_{lm}
\begin{pmatrix}
\varphi_{l\,1}(t)\chi_{lm}^-(\Omega_{d-1})\\
\\
i\phi_{l\, 1}(t)\chi_{lm}^-(\Omega_{d-1})
\end{pmatrix}
+D'_{lm}
\begin{pmatrix}
\varphi_{l\,2}(t)\chi_{lm}^-(\Omega_{d-1})\\
\\
-i\phi_{l\, 2}(t)\chi_{lm}^-(\Omega_{d-1})
\end{pmatrix},
\eqn
}
where $C(C')$ and $D(D')$ are arbitrary constants.

By examing the asymptotic behaviors of the general solution \eqref{eq:psi_even_gensol}, we could idenitfy the positive-/negative-frequency in-modes as the modes with the following asymptotic behaviors
\bea{
\Psi_{l+}&\sim e^{\left(\frac{d-1}{2}-iM\right)t}
\begin{pmatrix}
\chi_{lm}^+(\Omega_{d-1})\\
\\
-\chi_{lm}^+(\Omega_{d-1})
\end{pmatrix}
\text{ or }
e^{\left(\frac{d-1}{2}-iM\right)t}
\begin{pmatrix}
\chi_{lm}^-(\Omega_{d-1})\\
\\
-\chi_{lm}^-(\Omega_{d-1})
\end{pmatrix},\\
\Psi_{l-}&\sim e^{\left(\frac{d-1}{2}+iM\right)t}
\begin{pmatrix}
\chi_{lm}^+(\Omega_{d-1})\\
\\
\chi_{lm}^+(\Omega_{d-1})
\end{pmatrix}
\text{ or }
e^{\left(\frac{d-1}{2}+iM\right)t}
\begin{pmatrix}
\chi_{lm}^-(\Omega_{d-1})\\
\\
\chi_{lm}^-(\Omega_{d-1})
\end{pmatrix}
\quad \text{as } t\rightarrow-\infty,
\eqn
}
at the past infinity
while the positive/negative-frequency out-modes as the modes with the asymptotic behaviors
\bea{
\Psi_{l}^{\,\,+}&\sim e^{\left(-\frac{d-1}{2}-iM\right)t}
\begin{pmatrix}
\chi_{lm}^+(\Omega_{d-1})\\
\\
-\chi_{lm}^+(\Omega_{d-1})
\end{pmatrix}
\text{ or }
e^{\left(-\frac{d-1}{2}-iM\right)t}
\begin{pmatrix}
\chi_{lm}^-(\Omega_{d-1})\\
\\
-\chi_{lm}^-(\Omega_{d-1})
\end{pmatrix},\\
\Psi_{l}^{\,\,-}&\sim e^{\left(-\frac{d-1}{2}+iM\right)t}
\begin{pmatrix}
\chi_{lm}^+(\Omega_{d-1})\\
\\
\chi_{lm}^+(\Omega_{d-1})
\end{pmatrix}
\text{ or }
e^{\left(-\frac{d-1}{2}+iM\right)t}
\begin{pmatrix}
\chi_{lm}^-(\Omega_{d-1})\\
\\
\chi_{lm}^-(\Omega_{d-1})
\end{pmatrix}
\quad \text{as } t\rightarrow+\infty,
\eqn
}
at the future infinity.
In terms of these in-/out-modes, the Bogoliubov transformation \eqref{eq:bog_spinor} is expressed as
\bea{\label{eq:bogcoe_spin}
\Psi_{l\,+}&=\mu_l\,\Psi_l^{\,\, +}+\nu_l\,\Psi_l^{\,\, -},\\
\Psi_{l\,-}&=-\nu_l^*\,\Psi_l^{\,\, +}+\mu_l^*\,\Psi_l^{\,\, -},
\eqn
}
and we find the corresponding Bogoliubov coefficients to be
\bea{
\mu_{\,l}&=\frac{\Gamma(\frac{1}{2}-iM)^2}{\Gamma(1-l-\frac{d}{2}-iM)\Gamma(\frac{d}{2}+l-iM)},
\quad (l=0,1,\dots)\\
\nu_{\,l}&=\mp\frac{i\cos(l\pi+\frac{d}{2}\pi)}{\cosh(\pi M)},
\eqn
}
where $(-)$ sign is taken for the modes with $\chi^+_{lm}$ components while the $(+)$ sign is taken for the modes with $\chi^-_{lm}$ components. One can verify that these coefficients satisfy the relation
\be
|\mu_l|^2+|\nu_l|^2=1.
\ee
as required by the commutation rules.
The degeneracy for each mode $l$ is
\be
\mathcal{D}_{d-1}(l)=\frac{2^{\frac{d}{2}}(d+l-2)!}{l!\,(d-2)!},\quad \text{for even } d.
\ee

\subsection{Odd dimension $(d\geq 3)$}
In this case the dimension of the gamma matrices is $2^{(d-1)/2}$, same as the $d-1$ dimension representation. If we let $\{\tilde{\Gamma}^i\}$ be the set of $d-1$ matrices of dimension $2^{(d-1)/2}$ which satisfies the Clifford algebra \eqref{eq:clifford_algebra}, then the set of matrices
\be\label{eq:def_gamma_odd}
\gamma^0=
\begin{pmatrix}
i\mathbf{1} & 0\\
0 & -i\mathbf{1}    
\end{pmatrix},
\quad 
\gamma^j=\tilde{\Gamma}^j,
\quad (j=1,\dots,d-1)
\ee
satisfies the Dirac algebra \eqref{eq:id_gamma}. 
Using the representations of the gamma matrices \eqref{eq:def_gamma_odd} and \eqref{eq:comp_spinconnection}, the Dirac equation \eqref{eq:Dirac_general} becomes
\be\label{eq:dirac_odd_1}
\gamma^0\left(\partial_t+\frac{d-1}{2}\tanh(t)\right)\Psi+\frac{1}{\cosh(t)}\tilde{\slashed{\nabla}}\Psi+M\Psi=0,
\ee
where $\tilde{\slashed{\nabla}}$ is the Dirac operator on $S^{d-1}$. Instead of solving \eqref{eq:dirac_odd_1} directly, it is easier for us to act with the operator $\slashed{\nabla}-M$ on both sides and solve the following equation instead
\be\label{eq:dirac_odd_2}
-\left(\partial_t+\frac{d-1}{2}\tanh(t)\right)^2\Psi-\frac{\tanh(t)}{\cosh(t)}\gamma^0\tilde{\slashed{\nabla}}\Psi+\frac{1}{\cosh^2(t)}\tilde{\slashed{\nabla}}^2\Psi-M^2\Psi = 0.
\ee
The reason is that the operator $\gamma^0\tilde{\slashed{\nabla}}$ commutes with $\tilde{\slashed{\nabla}}^2$ while it is not true for $\gamma^0$ and $\tilde{\slashed{\nabla}}$ in \eqref{eq:dirac_odd_1}, which follows from 
\be
\gamma^0\tilde{\slashed{\nabla}}+\tilde{\slashed{\nabla}}\gamma^0=0.
\ee
If we consider the eigenfunctions of the Dirac operator $\tilde{\slashed{\nabla}}$ on $S^{d-1}$ for $d$ odd which satisfy \cite{Camporesi:1995fb}
\be\label{eq:odd_spin_harm1}
\tilde{\slashed{\nabla}}\chi^{-}_{lm}(\Omega_{d-1})=-i\left(l+\frac{d-1}{2}\right)\chi^{-}_{lm}(\Omega_{d-1}),\quad (l=0,1,\dots)
\ee
with degeneracies
\be
\mathcal{D}^{(-)}_{d-1}(l)=\frac{2^{(d-1)/2}(d+l-2)!}{l!(d-2)!}\quad \text{for odd } d,
\ee
then one can verify that the functions $\chi^+_{lm}\equiv \gamma^0\chi^-_{lm}$ are also the eigenfunctions of $\tilde{\slashed{\nabla}}$ with eigenvalues
\be
\tilde{\slashed{\nabla}}\chi^{+}_{lm}(\Omega_{d-1})=i\left(l+\frac{d-1}{2}\right)\chi^{+}_{lm}(\Omega_{d-1}),\quad (l=0,1,\dots).
\ee 
Using $\chi^{(\pm)}_{lm}$, we can construct the following functions
\bea{
\hat{\chi}_{lm}^-(\Omega_{d-1})&=\frac{1}{\sqrt{2}}\bigl[\chi^-_{lm}(\Omega_{d-1})+\chi^+_{lm}(\Omega_{d-1})\bigr], \eqn\\
\hat{\chi}_{lm}^+(\Omega_{d-1})&=\gamma^0\hat{\chi}_{lm}^-(\Omega_{d-1}), \eqn
}
which are the common eigenfunctions of $\gamma^0\tilde{\slashed{\nabla}}$ and $\tilde{\slashed{\nabla}}^2$ with eigenvalues
\bea{
\tilde{\slashed{\nabla}}^2\hat{\chi}_{lm}^{(\pm)}(\Omega_{d-1})&=-\left(l+\frac{d-1}{2}\right)^2\hat{\chi}_{lm}^{(\pm)}(\Omega_{d-1}), \eqn\\
\gamma^0\tilde{\slashed{\nabla}}\hat{\chi}_{lm}^{(\pm)}(\Omega_{d-1})&=\pm i\left(l+\frac{d-1}{2}\right)\hat{\chi}_{lm}^{(\pm)}(\Omega_{d-1}). \eqn
}
Now we can expand $\Psi$ in terms of those functions as
\be\label{eq:Psi_exp}
\Psi(t,\Omega_{d-1})=\sum_{l,m}\hat{\phi}_l(t)\hat{\chi}_{lm}^+(\Omega_{d-1})+
\hat{\varphi}_l(t)\hat{\chi}_{lm}^-(\Omega_{d-1}),
\ee
and substitute it into \eqref{eq:dirac_odd_2}. As a result, we find that $\hat{\phi}_l$ and $\hat{\varphi}_l$ satisfy the same equations \eqref{eq:dirac_even_phi} and \eqref{eq:dirac_even_varphi} for $\phi_l$ and $\varphi_l$ respectively. By further checking the Dirac equation \eqref{eq:dirac_odd_1}, one find that the general solution for $\Psi$ can be written as the ones satisfy
\bea{\label{eq:psi_odd_gensol}
\Psi=\sum_{l,m} &C_{lm}\left[\phi_{l\,1}(t)\hat{\chi}_{lm}^+(\Omega_{d-1})+\varphi_{l\,1}(t)\hat{\chi}_{lm}^-(\Omega_{d-1})\right]\\
&+
D_{lm}\left[\phi_{l\,2}(t)\hat{\chi}_{lm}^+(\Omega_{d-1})-\varphi_{l\,2}(t)\hat{\chi}_{lm}^-(\Omega_{d-1})\right],
\eqn
}
where $C$ and $D$ are arbitrary constants and $\phi_l$ and $\varphi_l$ are the same functions defined in the even $d$ case.
By examing the asymptotic behaviors of the general solution \eqref{eq:psi_odd_gensol}, we could identify the positive-/negative-frequency in-modes as those behave like
\bea{
\Psi_{l+}&\sim e^{(\frac{d-1}{2}-iM)t}\bigl[\hat{\chi}^+_{lm}(\Omega_{d-1})-i\hat{\chi}^-_{lm}(\Omega_{d-1})\bigr],\\
\Psi_{l-}&\sim e^{(\frac{d-1}{2}+iM)t}\bigl[\hat{\chi}^+_{lm}(\Omega_{d-1})+i\hat{\chi}^-_{lm}(\Omega_{d-1})\bigr]
\quad
\text{as } t\rightarrow -\infty,
\eqn
}
at the past infinity while the positive-/negative-frequency out-modes as those behave like
\bea{
\Psi_{l}^{\,\,+}&\sim e^{(-\frac{d-1}{2}-iM)t}\bigl[\hat{\chi}^+_{lm}(\Omega_{d-1})-i\hat{\chi}^-_{lm}(\Omega_{d-1})\bigr],\\
\Psi_{l}^{\,\,-}&\sim e^{(-\frac{d-1}{2}+iM)t}\bigl[\hat{\chi}^+_{lm}(\Omega_{d-1})+i\hat{\chi}^-_{lm}(\Omega_{d-1})\bigr]
\quad
\text{as } t\rightarrow +\infty,
\eqn
}
at the future infinity.
In terms of these in-/out-modes and the relation \eqref{eq:bogcoe_spin}, we find the corresponding Bogoliubov coefficients to be
\bea{
\mu_{\,l}&=\frac{\Gamma(\frac{1}{2}-iM)^2}{\Gamma(1-l-\frac{d}{2}-iM)\Gamma(\frac{d}{2}+l-iM)},
\quad (l=0,1,\dots)\\
\nu_{\,l}&=-\frac{i\cos(l\pi+\frac{d}{2}\pi)}{\cosh(\pi M)}=0
\quad \text{for odd } d,
\eqn
}
which has the same expression as in the even $d$ case. The degeneracy for each mode $l$ is 
\be
\mathcal{D}_{d-1}(l)=\frac{2^{(d-1)/2}(d+l-2)!}{l!(d-2)!}\quad \text{for odd } d.
\ee

\section{Particle production in even dimensions}
\label{sec:particle production}
Before proceeding to calculate the effective action, we could see there is a difference between even dimensions and odd dimensions. In \cite{Bousso:2001mw} it has been shown that the Bogoliubov coefficient $\nu_l$ vanishes when $d$ is odd in the massive scalar case. From our calculation, we see that it is also true in the massive spinor case. This implies that the in-vacuum and the out-vacuum are the same state in odd dimensions, so there is no production of either scalar or spinor particles. In contrast, there is always particle production in even dimensions for both scalar and spinor field. The particle production rate per spacetime volume $\mathcal{P}$ is related to the imaginary part of the effective action $W$ by
\be\label{eq:production rate}
\mathcal{P}=\lim_{V_d\rightarrow\infty}\frac{2}{V_d}\im W,
\ee
where $V_d$ is the spacetime volume of $dS_d$.

Using the Bogoliubov coefficients calculated in section \ref{sec:massive scalar} and \ref{sec:massive dirac} and the formula \eqref{eq:form_effact}, we obtain the effective action for a massive real scalar 
\bea{\label{eq:eff_sca_sum}
W_b=\frac{i}{2}\sum_{l=0}^\infty D_{(d-1)}(l)& \left[\ln\Gamma(1+i\mu)+\ln\Gamma(i\mu)-\ln\pi+\ln\sin(-l\pi-\tfrac{d-3}{2}\pi+i\mu\pi)\right.\\
&\left.
-\ln\Gamma\left(l+\tfrac{d-1}{2}+i\mu\right)
+\ln\Gamma\left(l+\tfrac{d-1}{2}-i\mu\right)\right],
\eqn
}
and the effective action for a massive Dirac spinor
\bea{\label{eq:eff_spi_sum}
W_f=-i\sum_{l=0}^\infty \mathcal{D}_{(d-1)}(l)& \left[2\ln\Gamma(\tfrac{1}{2}+i M)-\ln\pi+\ln\sin(-l\pi-\tfrac{d-2}{2}\pi+i M\pi)\right.\\
&\left.
-\ln\Gamma\left(l+\tfrac{d}{2}+i M\right)
+\ln\Gamma\left(l+\tfrac{d}{2}-i M\right)\right].
\eqn
}
In arriving at \eqref{eq:eff_sca_sum} and \eqref{eq:eff_spi_sum}, we have used the following identity for the gamma function
\be
\Gamma(1-z)\Gamma(z)=\frac{\pi}{\sin(\pi z)},\quad (z\notin\mathbb{Z}).
\ee
By using the integral representation for $\ln\Gamma(z)$ \cite{Gradshtein:2015}:
\be
\ln\Gamma(z)=\int^\infty_0\frac{ds}{s}\left[\frac{e^{-zs}-e^{-s}}{1-e^{-s}}+(z-1)e^{-s}\right],\quad (\re(z)>0),
\ee
we obtain that for odd $d$, the effective actions are
\bea{
W_b=\frac{1}{2}\sum_{l=0}^\infty D_{(d-1)}(l)\biggl[\int_0^\infty\frac{ds}{s}\frac{\sin(\mu s)}{\sinh(\frac{s}{2})}\left(e^{-\frac{s}{2}}-e^{(-\frac{d}{2}-l+1)s}\right)+\frac{(-1)^{\frac{d+1}{2}+l}-1}{2}\pi\biggr],
\eqn
}
\bea{
W_f=
\sum_{l=0}^\infty \mathcal{D}_{(d-1)}(l)\biggl[\int_0^\infty\frac{ds}{s}\frac{\sin(M s)}{\sinh(\frac{s}{2})}\left(e^{(-\frac{d}{2}-l+\frac{1}{2})s}-1\right)+\frac{(-1)^{\frac{d+1}{2}+l}+1}{2}\pi\biggr],
\eqn
}
while for even $d$ the effective actions are
\bea{\label{eq:im_scalar0}
W_b=\frac{1}{2}\sum_{l=0}^\infty D_{(d-1)}(l)&\biggl[\int_0^\infty\frac{ds}{s}\frac{\sin(\mu s)}{\sinh(\frac{s}{2})}\left(e^{-\frac{s}{2}}-e^{(-\frac{d}{2}-l+1)s}\right)
+\frac{(-1)^{\frac{d-2}{2}+l}}{2}\pi\\
&\quad +i\ln\coth(\pi\mu)\biggr],
\eqn
}
\bea{
W_f=\sum_{l=0}^\infty \mathcal{D}_{(d-1)}(l)&\biggl[\int_0^\infty\frac{ds}{s}\frac{\sin(M s)}{\sinh(\frac{s}{2})}\left(e^{(-\frac{d}{2}-l+\frac{1}{2})s}-1\right)+\frac{(-1)^{\frac{d-2}{2}+l}}{2}\pi\\
&\quad+i\ln\coth(\pi M)\biggr].
\eqn
}
Due to the infinite summation over the angular quantum number $l$, the scalar and spinor effective actions are divergent  in both even and odd dimensions. However, in odd dimensions the effective action is pure real while in even dimensions it has an divergent imaginary part besides the divergent real part:
\begin{align}
\label{eq:imW_b}
\im W_b&=\frac{1}{2}\sum_{l=0}^\infty D_{(d-1)}(l)\ln\coth(\pi\mu),\\
\label{eq:imW_f}
\im W_f&=\sum_{l=0}^\infty \mathcal{D}_{(d-1)}(l)\ln\coth(\pi M).
\end{align}
Since the divergence of the imaginary part is proportional to the summation of the degeneracies for both scalar and spinor field, we could regularize it by introducing a cut-off $N\gg 1$ on the angular quantum number $l$ \footnote{Since the summation for the real part of the effective action is more complicated, we will use a more covariant approach to regularize the effective action in section \ref{sec:effaction dim_reg}.}. To relate this cut-off $N$ to the cut-off $T$ on the time discussed in section \ref{sec:intro}, we use the method developed in \cite{Mottola:1984ar,Anderson:2013ila}. This method is based on the analysis of the real time particle creation process, which requires that we evolve system from a finite initial time $-T$ to a finite final time $T$ and the $T\to\infty$ limit is taken at the end. By examing the wave equantions \eqref{eq:KG-scalar}, \eqref{eq:dirac_even_phi} and \eqref{eq:dirac_even_varphi}, we see that the cut-off $N$ corresponds to a cut-off on the physical momentum
\be
k_{\text{phys}}\sim \frac{N}{\cosh(T)},
\ee 
at the time when the initial state is prepared. Altough it seems that the cut-off $N$ directly corresponds to a UV cut-off on the physical momentum, this is not true since we also need to take the $T\to\infty$ limit in the end. In fact if we demand that the cut-off for $k_{\text{phys}}$ is fixed at the finite initial time when the state is prepared, we need to change the cut-off $N$ accordingly when taking the limit $T\to\infty$. Specifically, the change of $T$ by $\delta T$ requires a change of $N$ by
\be\label{eq:cutoff_relation}
\delta N\approx N  \delta T, \text{ or } N\approx e^{T}.
\ee
Therfore, the divergences in the summations \eqref{eq:imW_b} and \eqref{eq:imW_f} result from the IR divergence of the spacetime volume. 

If we change the time cut-off $T$ by $\delta T$, then the spacetime volume changes by
\be
\delta V\approx \frac{2\pi^{\frac{d}{2}}d}{\Gamma(1+\frac{d}{2})}\cosh^{d-1}(T)\delta T.
\ee
In the meantime, the cut-off $N$ needs to be changed by $\delta N$ as we have argued before. This results a change in $\im W$ by
\begin{align}
\delta \im W_b &\approx \frac{2\ln\coth(\pi\mu)}{\Gamma(d-1)}N^{d-2}\delta N,\\
\delta \im W_f &\approx \frac{2^{\frac{d}{2}}\ln\coth(\pi M)}{\Gamma(d-1)}N^{d-2}\delta N.
\eqn
\end{align}
Using \eqref{eq:cutoff_relation} and the definition of $\mathcal{P}$ in \eqref{eq:production rate}, we find that the particle production rate for a massive real scalar is
\be
\mathcal{P}_b\approx\frac{2^{d-1}\Gamma(\frac{d}{2}+1)}{d\,\pi^{\frac{d}{2}}\Gamma(d-1)}\ln\coth(\pi \mu),
\quad (d \text{ even}),
\ee
and the particle production rate for a massive Dirac spinor is
\be
\mathcal{P}_f\approx\frac{2^{\frac{3d}{2}-1}\Gamma(\frac{d}{2}+1)}{d\,\pi^{\frac{d}{2}}\Gamma(d-1)}\ln\coth(\pi M),
\quad (d \text{ even}).
\ee
In the large-mass/weak-curvature limit ($M\gg H$), the particle production rate for both scalar and spinor fields goes like
\be
\mathcal{P}\sim e^{-M/T_H}.
\ee
where $T_H=H/(2\pi)$ is the Hawking-de Sitter temperature. This agrees with the results calculated in \cite{Akhmedov:2009ta,Akhmedov:2019esv} using the Green's function method.

\section{The vacuum amplitude and the ratio of determinants}
\label{sec:effaction dim_reg}
In section \ref{sec:particle production}, we find that the expressions for the effective action \eqref{eq:eff_sca_sum} and \eqref{eq:eff_spi_sum} are divergent due to the infinite summation over the angular quantum number $l$. From the structures of divergence, i.e., \eqref{eq:div_structure_even} for even $d$ and \eqref{eq:div_structure_odd} for odd $d$, we expect that the effective action contains a logarithmically divergent term in odd dimensions but no such term in even dimensions. Therefore, the finite piece of the effective action is unambiguous in even dimensions. In the language of dS/CFT, we expect that the coefficient of the logarithmically divergent term in odd dimensions might be connected to the conformal anomaly present in the boundary CFT\cite{Deser:1993yx,Henningson:1998gx,Mazur:2001aa,Nojiri:2001mf}.

In this section, we compute the finite term of the effective action in even dimensions and the coefficient of the logarithmically divergent term of the effective action in odd dimensions. We follow the method developed in \cite{Diaz:2007an} and use dimensional regularization\footnote{In the appendix, we show that the similar result can be obtained using another regularization method.} to regularize the summation over $l$ in \eqref{eq:eff_sca_sum} and \eqref{eq:eff_spi_sum}. In dimensional regularization, the logarithmically divergent term corresponds to the pole in $\epsilon$ \cite{Deser:1993yx}, where we set the dimension $d=\text{integer}-\epsilon$. Along the way, we show that the regularized vacuum amplitude $\Z$ in $dS_d$ has the same expression as the ratio of the functional determinants associated with different quantizations in $AdS_d$. The calculation of such ratio of the determinants have appeared in the study of double-trace deformation in AdS/CFT correspondence\cite{Gubser:2002zh,Gubser:2002vv,Diaz:2007an,Allais:2010qq,Aros:2011iz}. 

\subsection{Real massive scalar}
The vacuum amplitude $\Z^b$ is related to the effective action $W_b$ through the expression $\log\Z^b=iW_b$. Therefore, we have
\bea{
\log\Z^b=-\frac{1}{2}\sum_{l=0}^\infty D_{(d-1)}(l)& \left[\ln\Gamma(1+i\mu)+\ln\Gamma(i\mu)-\ln\pi+\ln\sin(-l\pi-\tfrac{d-3}{2}\pi+i\mu\pi)\right.\\
&\left.
-\ln\Gamma\left(l+\tfrac{d-1}{2}+i\mu\right)
+\ln\Gamma\left(l+\tfrac{d-1}{2}-i\mu\right)\right].
\eqn
}
For reasons that will become clear later, we define $\nu=i\mu=i\sqrt{M^2-\frac{(d-1)^2}{4}}$ and consider the derivative of $\log\Z^b$ with respect to $\nu$:
\be\label{eq:dlogZb}
\frac{1}{2\nu}\frac{\partial}{\partial\nu}\log\Z^b
=\frac{1}{4\nu}\sum_{l=0}^\infty D_{(d-1)}(l)\biggl[
\psi(l+\tfrac{d-1}{2}+\nu)
+\psi(l+\tfrac{d-1}{2}-\nu)
\biggr],
\ee
where $\psi(z)$ is the digamma function. In \eqref{eq:dlogZb} we have neglected all the terms that are proportional to $\sum_{l=0}^\infty D_{(d-1)}(l)$. The reason is that $\sum_{l=0}^\infty D_{(d-1)}(l)=0$ under dimensional regularization \cite{Diaz:2007an}. We briefly review the argument here. The degeneracy $D_{(d-1)}(l)$ can be rewritten as
\be
D_{(d-1)}(l)=\frac{2l+d-2}{d-2}\frac{(d-2)_l}{l!},
\ee
where $(a)_l=\Gamma(a+l)/\Gamma(a)$ is the Pochhammer symbol. Now using the following expansion for $(1-x)^a$:
\be
(1-x)^a=\sum^\infty_{l=0}\frac{(-a)_l}{l!}x^l,
\ee
we have
\be
\sum_{l=0}^\infty D_{(d-1)}(l)=2(1-1)^{-(d-1)}+(1-1)^{-(d-2)},
\ee
which is 0 for $d<1$. As the summation \eqref{eq:dlogZb} is also convergent when $d<1$, we can analytically continue the result from this region. To proceed, we use the integral representation for $\psi(z)$:
\be
\psi(z)=\int_0^\infty ds\,\left(\frac{e^{-s}}{s}-\frac{e^{-sz}}{1-e^{-s}}\right),
\ee
and perform the summation over $l$. The remaining integral over $s$ can be done analytically and the final expression is
\be\label{eq:det_sca_2}
\frac{1}{2\nu}\frac{\partial}{\partial\nu}\log\Z^b
=-\frac{1}{2}\Gamma(1-d)
\left[
\frac{\Gamma(\nu+\frac{d-1}{2})}{\Gamma(1+\nu-\frac{d-1}{2})}
-\frac{\Gamma(-\nu+\frac{d-1}{2})}{\Gamma(1-\nu-\frac{d-1}{2})}
\right].
\ee

On the other hand, the Euclidean one-loop effective action for a massive real scalar field in $AdS_d$ is
\be
Z_b^{\pm}=Z^b_{class}\cdot\left[{\det}_{\pm}(-\nabla^2+m^2)\right]^{-\frac{1}{2}}.
\ee
where $Z^b_{class}$ is the classical partition function and $\pm$ refers to the bulk quantization corresponding to the bounary operator with dimension $\D_{\pm}$ defined by
\be
\D_{\pm}=\frac{d-1}{2}\pm \nu',\quad \nu'=\sqrt{\frac{(d-1)^2}{4}+m^2}.
\ee
Instead of $\left[{\det}_{\pm}(-\nabla^2+m^2)\right]^{-\frac{1}{2}}$, it's much easier to calculate:
\be
\frac{\partial}{\partial m^2}\log\left[{\det}_{\pm}(-\nabla^2+m^2)\right]^{-\frac{1}{2}}=-\frac{1}{2}\int dr dx^{d-1}\sqrt{g}\,G^b_{\D_\pm}(r,x;r,x),
\ee
where $G^b_{\D_\pm}$ is the propagator for the scalar field. Using dimensional regularization, we have \cite{Allais:2010qq}
\be
G^b_{\D_\pm}(r,x;r,x)=(4\pi)^{-\frac{d}{2}}\Gamma(1-\tfrac{d}{2})\frac{\Gamma(\pm\nu'+\frac{d-1}{2})}{\Gamma(1\pm\nu'-\frac{d-1}{2})},
\ee
and the spacetime volume to be
\be\label{eq:VAdSd_reg}
V_d=\pi^{\frac{d-1}{2}}\Gamma(-\tfrac{d-1}{2}).
\ee
Thus, under dimensional regularization \cite{Diaz:2007an} we obtain
\bea{\label{eq:det_sca_1}
&\frac{\partial}{\partial m^2}\log\frac{\left[\det_+(-\nabla^2+m^2)\right]^{-\frac{1}{2}}}{\left[\det_-(-\nabla^2+m^2)\right]^{-\frac{1}{2}}}\\
=&-\frac{1}{2}\Gamma(1-d)
\left[
\frac{\Gamma(\nu'+\frac{d-1}{2})}{\Gamma(1+\nu'-\frac{d-1}{2})}
-\frac{\Gamma(-\nu'+\frac{d-1}{2})}{\Gamma(1-\nu'-\frac{d-1}{2})}
\right].
\eqn
}

If we identify $\nu'$ with $\nu$, then we can establish the equality
\be
\frac{1}{2\nu}\frac{\partial}{\partial\nu}\log\Z^b=
\frac{\partial}{\partial m^2}\log\frac{\left[\det_+(-\nabla^2+m^2)\right]^{-\frac{1}{2}}}{\left[\det_-(-\nabla^2+m^2)\right]^{-\frac{1}{2}}}
=\frac{1}{2\nu'}\frac{\partial}{\partial\nu'}\log\frac{Z_b^+}{Z_b^-},
\ee
under dimensional regularization.

Although there appears to be a pole for all physical dimension $d$ in \eqref{eq:det_sca_2}, the expression actually only has a pole in $d$ when $d$ is odd:
\bea{\label{eq:det_sca_f}
&-\frac{1}{2}\Gamma(1-d)
\left[
\frac{\Gamma(\nu+\frac{d-1}{2})}{\Gamma(1+\nu-\frac{d-1}{2})}
-\frac{\Gamma(-\nu+\frac{d-1}{2})}{\Gamma(1-\nu-\frac{d-1}{2})}
\right]\\
=&\frac{\sin(\pi\nu)}{2\cos(\frac{d\pi}{2})}\frac{\Gamma(\nu+\tfrac{d-1}{2})\Gamma(-\nu+\tfrac{d-1}{2})}{\Gamma(d)}.
\eqn
}
Now we go to the physical dimension by letting $d\rightarrow d-\epsilon$.
\subparagraph{$d$ even}
In this case, \eqref{eq:det_sca_f} is finite:
\be\label{eq:sca_d_even}
\frac{1}{2\nu}\frac{\partial}{\partial\nu}\log\Z^b
=\frac{\pi}{2\nu}\frac{(-1)^{\frac{d}{2}+1}}{\Gamma(d)}\frac{\Gamma(\nu+\tfrac{d-1}{2})\Gamma(-\nu+\tfrac{d-1}{2})}{\Gamma(\nu)\Gamma(-\nu)}.
\ee
In the language of dS/CFT, the boundary CFT also has no conformal anomaly and the finite term in the partition function of the CFT is well defined. The result \eqref{eq:sca_d_even} gives the difference of this finite term in the UV and IR CFT \cite{Gubser:2002zh,Gubser:2002vv,Diaz:2007an}.

\subparagraph{$d$ odd}
In this case, \eqref{eq:det_sca_f} has a pole in $\epsilon$
\be\label{eq:sca_d_odd}
\frac{1}{2\nu}\frac{\partial}{\partial\nu}\log\Z^b
=\frac{1}{\epsilon}\frac{(-1)^{\frac{d+1}{2}}}{\nu\Gamma(d)}\frac{\Gamma(\nu+\tfrac{d-1}{2})\Gamma(-\nu+\tfrac{d-1}{2})}{\Gamma(\nu)\Gamma(-\nu)}+\mathcal{O}(1),
\ee
which signals the logarthmic divergence. In the language of dS/CFT, the boundary CFT has conformal anomaly in this case. The change of the conformal anomaly due to the double-trace deformation can be computed from the change of the central charge between the UV and IR CFT. The residue at the pole in \eqref{eq:sca_d_odd} reproduces this change of the central charge on the boundary CFT\cite{Gubser:2002vv} up to a constanct prefactor.

\subsection{Massive Dirac spinor}
For a massive Dirac spinor, we have
\bea{
\log\Z^f=\sum_{l=0}^\infty \mathcal{D}_{(d-1)}(l)& \left[2\ln\Gamma(\tfrac{1}{2}+i M)-\ln\pi+\ln\sin(-l\pi-\tfrac{d-2}{2}\pi+i M\pi)\right.\\
&\left.
-\ln\Gamma\left(l+\tfrac{d}{2}+i M\right)
+\ln\Gamma\left(l+\tfrac{d}{2}-i M\right)\right].
\eqn
}
Similar to the scalar case, we denote $\nu=iM$ and consider the derivative of $\log \Z^f$ with respect to $\nu$:
\be\label{eq:dlogZf}
\frac{\partial}{\partial\nu}\log\Z^f
=-\sum_{l=0}^\infty \mathcal{D}_{(d-1)}(l)\biggl[
\psi(l+\tfrac{d}{2}+\nu)
+\psi(l+\tfrac{d}{2}-\nu)
\biggr].
\ee
We recall that the degeneracies can be written as
\be
\mathcal{D}_{(d-1)}(l)=
\dim\gamma^d\,\frac{(d-1)_l}{l!}.
\ee
where $\dim\gamma^d$ is the dimmension of the gamma matrices in $d$-dimensional spacetime. Again the $l-$independent terms in \eqref{eq:dlogZf} is neglected because $\sum_{l=0}^\infty \mathcal{D}_{(d-1)}(l)=0$ using dimensional regularization. Following the same method used in the scalar case, we find that the final expression is
\be\label{eq:det_spi_2}
\frac{\partial}{\partial\nu}\log\Z^f
=\dim\gamma^d\,\Gamma(1-d)
\left[
\frac{\Gamma(\nu+\frac{d}{2})}{\Gamma(1+\nu-\frac{d}{2})}
+\frac{\Gamma(-\nu+\frac{d}{2})}{\Gamma(1-\nu-\frac{d}{2})}
\right].
\ee

On the other hand, the Euclidean one-loop effective action for a massive Dirac spinor field in $AdS_d$ is:
\be
Z_f^{\pm}=Z^f_{class}\cdot\left[{\det}_{\pm}(\slashed{\nabla}+m)\right],
\ee
where $Z^f_{class}$ is the classical partition function and $\pm$ refers to the bulk quantization corresponding to the boundary operator with dimension $\D_{\pm}$ defined by
\be
\D_{\pm}=\frac{d-1}{2}\pm \nu',\quad \nu' = m.
\ee
As in the scalar case, it is much easier to compute
\be
\frac{\partial}{\partial m}\log\left[{\det}_{\pm}(\slashed{\nabla}+m)\right]=-\int drdx^{d-1}\sqrt{g}\,\Tr\left[G^f_{\D_{\pm}}(r,x;r,x)\right],
\ee
where $G^f_{\D_{\pm}}$ is the propagator for the spinor field. Using dimensional regularization, we have \cite{Allais:2010qq,Aros:2010ng}
\be
\Tr\left[G^f_{\D_{\pm}}(r,x;r,x)\right]=\mp\dim\gamma^d\,(4\pi)^{-\frac{d}{2}}\Gamma(1-\tfrac{d}{2})\frac{\Gamma(\frac{d}{2}\pm\nu')}{\Gamma(1-\frac{d}{2}\pm\nu')},
\ee 
Multiplying the regularized volume for spacetime \eqref{eq:VAdSd_reg}, we obtain
\bea{\label{eq:det_spi_1}
&\frac{\partial}{\partial m}\log\frac{\left[\det_+(\slashed{\nabla}+m)\right]}{\left[\det_-(\slashed{\nabla}+m)\right]}\\
=&
\dim\gamma^d\,\Gamma(1-d)
\left[
\frac{\Gamma(\nu'+\frac{d}{2})}{\Gamma(1+\nu'-\frac{d}{2})}
+\frac{\Gamma(-\nu'+\frac{d}{2})}{\Gamma(1-\nu'-\frac{d}{2})}
\right].
\eqn
}

If we identify $\nu'$ with $\nu$, we can establish the equality
\be
\frac{\partial}{\partial\nu}\log\Z^f
=\frac{\partial}{\partial m}\log\frac{\left[\det_+(\slashed{\nabla}+M)\right]}{\left[\det_-(\slashed{\nabla}+M)\right]}
=\frac{\partial}{\partial \nu'}\log\frac{Z^+_f}{Z^-_f},
\ee
under dimensional regularization.
Similar to the scalar case, the expression \eqref{eq:det_spi_2} only has pole in $d$ when $d$ is odd
\bea{\label{eq:det_spi_f}
&\dim\gamma^d\,\Gamma(1-d)
\left[
\frac{\Gamma(\nu+\frac{d}{2})}{\Gamma(1+\nu-\frac{d}{2})}
-\frac{\Gamma(-\nu+\frac{d}{2})}{\Gamma(1-\nu-\frac{d}{2})}
\right]\\
=&\dim\gamma^d\,\frac{\cos(\pi\nu)}{\cos(\frac{d\pi}{2})}\frac{\Gamma(\nu+\tfrac{d}{2})\Gamma(-\nu+\tfrac{d}{2})}{\Gamma(d)}.
\eqn
}

Now we go to physical dimension by letting $d\rightarrow d-\epsilon$.
\subparagraph{$d$ even}
In this case, \eqref{eq:det_spi_f} is finite:
\be\label{eq:spi_even_d}
\frac{\partial}{\partial\nu}\log\Z^f=\dim\gamma^d\,\frac{\pi(-1)^{\frac{d}{2}}}{\Gamma(d)}\frac{\Gamma(\nu+\tfrac{d}{2})\Gamma(-\nu+\tfrac{d}{2})}{\Gamma(\nu+\tfrac{1}{2})\Gamma(-\nu+\tfrac{1}{2})}.
\ee
As in the scalar case, the boundary CFT has no anomaly and the result \eqref{eq:spi_even_d} computes the difference of the finite term in UV and IR CFT \cite{Aros:2011iz}. 

\subparagraph{$d$ odd}
In this case, \eqref{eq:det_spi_f} has a pole in $\epsilon$:
\be
\frac{\partial}{\partial\nu}\log\Z^f=\frac{1}{\epsilon}\dim\gamma^d\,\frac{(-1)^{\frac{d-1}{2}}}{2\Gamma(d)}\frac{\Gamma(\nu+\tfrac{d}{2})\Gamma(-\nu+\tfrac{d}{2})}{\Gamma(\nu+\tfrac{1}{2})\Gamma(-\nu+\tfrac{1}{2})}
+\mathcal{O}(1).
\ee
As in the scalar case, the boundary CFT has conformal anomaly which can be computed from its central charge. Up to a constanct prefactor, the residue at the pole reproduces the change of the central charge in the UV and IR CFT connected by the RG flow due to the double-trace deformation on the boundary CFT\cite{Allais:2010qq,Aros:2011iz}.

\section{Conclusion}
In this paper we have used in-/out-state formalism to calculate the effective action for both a real scalar field and a Dirac spinor field in the global patch of dS space in any dimension. It has been known for a long time \cite{Bousso:2001mw} that there is no imaginary contribution to the effective action of a scalar field in the odd-dimensional dS space. In this paper we have shown that it is also true for the effective action of a spinor field in the odd-dimensional dS space. In \cite{Akhmedov:2019esv} the authors have given a heuristic argument for why there is no imaginary contribution for the scalar field in odd dimensions. We think this argument can be adapted to the spinor case as well. In even dimensions, there is an imaginary part in the effective action in both scalar and spinor field cases. Such imaginary part signals the event of  particle production. In both cases, we have calculated the corresponding particle production rate and we have found that in the large-mass/weak-curvature limit both rates approach $e^{-M/T_H}$ where $T_H$ is the Hawking-de Sitter temperature.

Using dimensional regularization, we have extracted the finite term of the effective action in even dimensions and the coeffient of the logarithmically divergent term in odd dimensions. We also have shown that the regularized in-out vacuum amplitude $\Z$ in global $dS_d$ has the same expression as the ratio of the functional determinants associated with different quantizations in $AdS_d$ upon identification of certain parameters in the two theories. It is intriguing that there is a relation between the vacuum amplitude in $dS_d$ and the ratio of determinants in $AdS_d$. We don't know if it is just a coincidence or there is a deeper connection underlying. Nevertheless, we want to point out that the summations in \eqref{eq:dlogZb} and \eqref{eq:dlogZf} have appeared exactly in the dual $CFT_{d-1}$ calculation \cite{Gubser:2002vv,Diaz:2007an,Allais:2010qq,Aros:2011iz} in the study of double trace deformation. In the $CFT_{d-1}$ computation for odd $d$, the coefficient of the logarithmic divergence is related to the change of the central charge from the UV fixed point to the IR fixed point\cite{Gubser:2002vv,Allais:2010qq}. The CFT at the two fixed points correspond to the two different quantizations in the bulk AdS space. In the language of dS/CFT \cite{Strominger:2001pn}, the in-/out-modes in our calculation also correspond to different quantizations in the dual CFT. It could be possible that the in-out vacuum amplitude in dS space is related to the double-trace deformation on the boundary CFT as the time evolution in the bulk corresponds to the RG flows in the dual CFT in the context of dS/CFT\cite{Das:2013qea}. It would be interesting if we could find the exact connection between the two.

\vspace{0.75cm}
{\bf Acknowledgments } 
We want to thank Juan Maldacena for helpful discussions and
valuable suggestions on the manuscript. We also want to thank Fedor Popov and Junyi Zhang for interesting discussions.

\appendix
\section{Regularization with Laplacian}
In this appendix, we show that the results in section \ref{sec:effaction dim_reg} could be obtained by other regularization method. Specifically, we regularize the summation over $l$ using the exponential suppression factor $\exp(-\lambda_l\, \varepsilon)$ in the scalar field case. The $\lambda_l=l(l+d-2)$ is the eigenvalue of the sphere Laplacian for the angular quantum number $l$. We perform the calculation for the case of $d=2,3,4,5$. We obtain the same finite terms as those calculated in section \ref{sec:effaction dim_reg} for $d=2,4$. However, for $d=3, 5$, the results differ by a finite term which could be obtained from the summation of the degeneracies $D_{(d-1)}(l)$. We think this mismatch might have resulted from a different choice of counterterms under the two regularization schemes.

We closely follow the method developed in \cite{Monin:2016bwf}. The following asymptotic expansion \cite{Monin:2016bwf} is crucial for our calculation:
\bea{\label{eq:lapreg_asy}
\sum_{l=1}^\infty l^{-s}e^{-l(l+q)t}\underset{t\rightarrow 0}{=}&
\frac{t^{-\frac{1-s}{2}}}{2}\biggl[\Gamma\left(\frac{1}{2}-\frac{s}{2}\right)-q\Gamma\left(1-\frac{s}{2}\right)t^\frac{1}{2}+
\frac{q^2}{2!}\Gamma\left(\frac{3}{2}-\frac{s}{2}\right)t\biggr.\\
\qquad&\biggl.
-\frac{q^3}{3!}\Gamma\left(2-\frac{s}{2}\right)t^\frac{3}{2}
+\frac{q^4}{4!}\Gamma\left(\frac{5}{2}-\frac{s}{2}\right)t^2+\mathcal{O}(t^2)
\biggr]
+\zeta(s).
\eqn
}

\paragraph{d=2}
We first look at the summation of the degeneracies using \eqref{eq:lapreg_asy}, which is
\be
1+\sum_{l=1}^\infty 2\,e^{-l^2\varepsilon}
\underset{\varepsilon\rightarrow 0}{=}\sqrt{\pi}\varepsilon^{-\frac{1}{2}}.
\ee
So the summation of the $l$-independent terms do not contain either $\log\varepsilon$ term or a finite term. The summation of the remaining terms can be evaluated as
\bea{
\frac{1}{2\nu}\frac{\partial}{\partial\nu}\log\Z^b
=&\frac{1}{4\nu}\biggl[\psi(\tfrac{1}{2}+\nu)
+\psi(\tfrac{1}{2}-\nu)\biggr] 
+\frac{1}{2\nu}\sum_{l=1}^\infty\biggl[
\psi(l+\tfrac{1}{2}+\nu)
+\psi(l+\tfrac{1}{2}-\nu)
\biggr]e^{-l^2\varepsilon}\\
=&\frac{1}{\nu}\sum_{l=1}^\infty\log l\,e^{-l^2\varepsilon}
+\frac{1}{2\nu}\sum_{l=1}^\infty\biggl[
\psi(l+\tfrac{1}{2}+\nu)
+\psi(l+\tfrac{1}{2}-\nu)
-2\log l
\biggr]\\
&
+\frac{1}{4\nu}\biggl[\psi(\tfrac{1}{2}+\nu)
+\psi(\tfrac{1}{2}-\nu)\biggr]\\
=&\frac{1}{2\nu}\log(2\pi)-\frac{1}{2\nu}\biggl[\log(2\pi)+\pi\nu\tan(\pi\nu)\biggr]\\
=&-\frac{\pi}{2}\tan(\pi\nu),
\eqn
}
where we have only kept the relavent finite term and the $\log\varepsilon$ term from the summation. To arrive at the result, we have used \eqref{eq:lapreg_asy} to compute the first summation on the second line, while the second summation is convergent (goes like $l^{-2}$ asymptotically) and can be calculated analytically. The result agrees with the finite term obtained using the dimensional regularization.

\paragraph{d=4}
The summation of the degeneracies is
\be
1+\sum_{l=1}^\infty (l+1)^2 e^{-l(l+2)\varepsilon}
\underset{\varepsilon\rightarrow 0}{=}
\frac{\sqrt{\pi}}{4}(\varepsilon^{-\frac{3}{2}}+\varepsilon^{-\frac{1}{2}}).
\ee
As in $d=2$ case, we can neglect the summation of $l$-independent terms. The remaining summation is
\bea{
&\frac{1}{2\nu}\frac{\partial}{\partial\nu}\log\Z^b\\
=&
\frac{1}{4\nu}\biggl[\psi(\tfrac{3}{2}+\nu)
+\psi(\tfrac{3}{2}-\nu)\biggr]\\
&+
\frac{1}{4\nu}\sum_{l=1}^\infty (l+1)^2\biggl[
\psi(l+\tfrac{3}{2}+\nu)
+\psi(l+\tfrac{3}{2}-\nu)
\biggr]e^{-l(l+2)\varepsilon}\\
=&\frac{1}{4\nu}\sum_{l=1}^\infty f(l)\,e^{-l(l+2)\varepsilon}
+\frac{1}{4\nu}\sum_{l=1}^\infty\biggl\{
(l+1)^2\biggl[
\psi(l+\tfrac{3}{2}+\nu)
+\psi(l+\tfrac{3}{2}-\nu)\biggr]-f(l)
\biggr\}\\
&+\frac{1}{4\nu}\biggl[\psi(\tfrac{3}{2}+\nu)
+\psi(\tfrac{3}{2}-\nu)\biggr]\\
=&\frac{1}{96\nu}\biggl[-75+16\gamma+36\nu^2+96\log\mathcal{G}+24\log(2\pi)-48\zeta'(-2)\biggr]\\
&+\frac{1}{96\nu}\biggl[75-16\gamma-36\nu^2-96\log\mathcal{G}-24\log(2\pi)-\frac{12}{\pi^2}\zeta(3)+2\pi\nu(1-4\nu^2)\tan(\pi\nu)\biggr]\\
=&-\frac{\pi}{48}(2\nu-1)(2\nu+1)\tan(\pi\nu),
\eqn
}
where $\mathcal{G}$ is Glaisher's constant and $\zeta(s)$ is the Riemann zeta function. We have only kept $\log\varepsilon$ and the finite term at the end. The function $f(l)$ equals
\be
f(l)=2l^2\log l+2l(1+2\log l)+\frac{1}{12}(37-12\nu^2+24\log l)+\frac{2}{3l}.
\ee
Again, the result agrees with the one obtained using the dimensional regularization.


\paragraph{d=3}
In this case, we expect there to be $\log\varepsilon$ term and we want to compute its coefficient. The summation of the degeneracies is
\be
1+\sum_{l=1}^\infty (2l+1) e^{-l(l+1)\varepsilon}
\underset{\varepsilon\rightarrow 0}{=}
\varepsilon^{-1}+\frac{1}{3}.
\ee
Thus, the summation of the $l$-independent terms will not contribute to the $\log\varepsilon$ term. The remaining summation is
\bea{\label{eq:lapreg_sc_d3}
&\frac{1}{2\nu}\frac{\partial}{\partial\nu}\log\Z^b\\
=&
\frac{1}{4\nu}\biggl[\psi(1+\nu)
+\psi(1-\nu)\biggr]\\
&+
\frac{1}{4\nu}\sum_{l=1}^\infty (2l+1)\biggl[
\psi(l+1+\nu)
+\psi(l+1-\nu)
\biggr]e^{-l(l+1)\varepsilon}\\
=&\frac{1}{4\nu}\sum_{l=1}^\infty f(l)\,e^{-l(l+1)\varepsilon}
+\frac{1}{4\nu}\sum_{l=1}^\infty\biggl\{
(2l+1)\biggl[
\psi(l+1+\nu)
+\psi(l+1-\nu)\biggr]-f(l)
\biggr\}\\
&+\frac{1}{4\nu}\biggl[\psi(1+\nu)
+\psi(1-\nu)\biggr],
\eqn
}
where $f(l)$ is given by
\be
f(l)=4l\log l+2(1+\log l)-\frac{2(3\nu^2-1)}{3l}.
\ee
As the second summation in \eqref{eq:lapreg_sc_d3} is convergent, the $\log\varepsilon$ term can only come from the first summation. We have
\be\label{eq:lapreg_sc_d3_c}
\frac{1}{2\nu}\frac{\partial}{\partial\nu}\log\Z^b\biggr|_{\log\varepsilon}=
\frac{1}{4\nu}\biggl(\nu^2-\frac{1}{3}\biggr)\log\varepsilon.
\ee
The dimensional regularization $(d=3-\epsilon)$ result is
\be
\frac{1}{2\nu}\frac{\partial}{\partial\nu}\log\Z^b\biggr|_{\frac{1}{\epsilon}}=-\frac{1}{2\nu}\nu^2\epsilon^{-1}.
\ee
The constant term in the parentheses of \eqref{eq:lapreg_sc_d3_c} equals
\be
\sum_{l=0}^\infty(2l+1)e^{-l(l+1)\varepsilon}\biggr|_{finite}=\frac{1}{3}.
\ee

\paragraph{d=5}
Again we want to calculate the coefficient of the $\log\epsilon$ term. The summation of the degeneracies is
\be
1+\sum_{l=1}^\infty \frac{(l+1)(l+2)(2l+3)}{6} e^{-l(l+3)\varepsilon}
\underset{\varepsilon\rightarrow 0}{=}
\frac{1}{6}\varepsilon^{-2}+\frac{1}{3}\varepsilon^{-1}+\frac{29}{90}.
\ee
So the summation of the $l$-independent terms will not contribute to $\log\epsilon$ term. The remaining summation is
\bea{\label{eq:lapreg_sc_d5}
\frac{1}{2\nu}\frac{\partial}{\partial\nu}\log\Z^b
=&
\frac{1}{4\nu}\biggl[\psi(2+\nu)
+\psi(2-\nu)\biggr]\\
&+
\frac{1}{4\nu}\sum_{l=1}^\infty \frac{(l+1)(l+2)(2l+3)}{6}\biggl[
\psi(l+2+\nu)
+\psi(l+2-\nu)
\biggr]e^{-l(l+3)\varepsilon}\\
=&\frac{1}{4\nu}\sum_{l=1}^\infty f(l)\,e^{-l(l+3)\varepsilon}
+\frac{1}{4\nu}\sum_{l=1}^\infty\biggl\{
\frac{(l+1)(l+2)(2l+3)}{6}\biggl[
\psi(l+2+\nu)\\
&+\psi(l+2-\nu)\biggr]
-f(l)
\biggr\}
+\frac{1}{4\nu}\biggl[\psi(2+\nu)
+\psi(2-\nu)\biggr].
\eqn
}
Again, only the first summation in \eqref{eq:lapreg_sc_d5} can contribute to $\log\epsilon$ as the other summation is convergent. Thus, we get
\be\label{eq:lapreg_sc_d5_c}
\frac{1}{2\nu}\frac{\partial}{\partial\nu}\log\Z^b\biggr|_{\log\varepsilon}=
\frac{1}{4\nu}\biggl(\frac{\nu^2(\nu-1)(\nu+1)}{12}-\frac{29}{90}\biggr)\log\varepsilon,
\ee
while the dimensional regularization $(d=5-\epsilon)$ result is
\be
\frac{1}{2\nu}\frac{\partial}{\partial\nu}\log\Z^b\biggr|_{\frac{1}{\epsilon}}=-\frac{1}{2\nu}\frac{\nu^2(\nu-1)(\nu+1)}{12}\epsilon^{-1}.
\ee
Similar to the $d=3$ case, the constant term in the parentheses of \eqref{eq:lapreg_sc_d5_c} equals
\be
\sum_{l=0}^\infty\frac{(l+1)(l+2)(2l+3)}{6}e^{-l(l+3)\varepsilon}\biggr|_{finite}=\frac{29}{90}.
\ee

\bibliographystyle{JHEP}
\bibliography{dSInOutRef}
\end{document}